\documentclass[Physsubmission, Phys]{SciPost}
\hypersetup{
    colorlinks,
    linkcolor={red!50!black},
    citecolor={blue!50!black},
    urlcolor={blue!80!black}
}

\usepackage[bitstream-charter]{mathdesign}

\DeclareSymbolFont{usualmathcal}{OMS}{cmsy}{m}{n}
\DeclareSymbolFontAlphabet{\mathcal}{usualmathcal}

\begin{document}

\begin{center}{\Large \textbf{
Hybrid high-energy/collinear factorization in a heavy-light dijets system reaction\\
}}\end{center}

% TODO: write the author list here. Use initials + surname format.
% Separate subsequent authors by a comma, omit comma at the end of the list.
% Mark the corresponding author with a superscript *.
\begin{center}
A. D. Bolognino\textsuperscript{1,2},
F. G. Celiberto\textsuperscript{3,4,5},
M. Fucilla\textsuperscript{1,2 $\star$},
D.Yu. Ivanov \textsuperscript{6},
A. Papa\textsuperscript{1,2},
\end{center}

\begin{center}
{\bf 1} Dipartimento di Fisica, Università della Calabria, \\ I-87036 Arcavacata di Rende, Cosenza, Italy
\\
{\bf 2} Istituto Nazionale di Fisica Nucleare, Gruppo collegato di Cosenza, \\ I-87036 Arcavacata di Rende, Cosenza, Italy
\\
{\bf 3} European Centre for Theoretical Studies in Nuclear Physics and Related Areas (ECT*), \\ I-38123 Villazzano, Trento, Italy
\\
{\bf 4} Fondazione Bruno Kessler (FBK), I-38123 Povo, Trento, Italy
\\
{\bf 5} INFN-TIFPA Trento Institute of Fundamental Physics and Applications, \\ I-38123 Povo, Trento, Italy
\\
{\bf 6} Sobolev Institute of Mathematics, 630090 Novosibirsk, Russia
\\
* michael.fucilla@unical.it
\end{center}

\begin{center}
\today
\end{center}

% For convenience during refereeing (optional),
% you can turn on line numbers by uncommenting the next line:
%\linenumbers

\definecolor{palegray}{gray}{0.95}
\begin{center}
\colorbox{palegray}{
  \begin{tabular}{rr}
  \begin{minipage}{0.1\textwidth}
    \includegraphics[width=22mm]{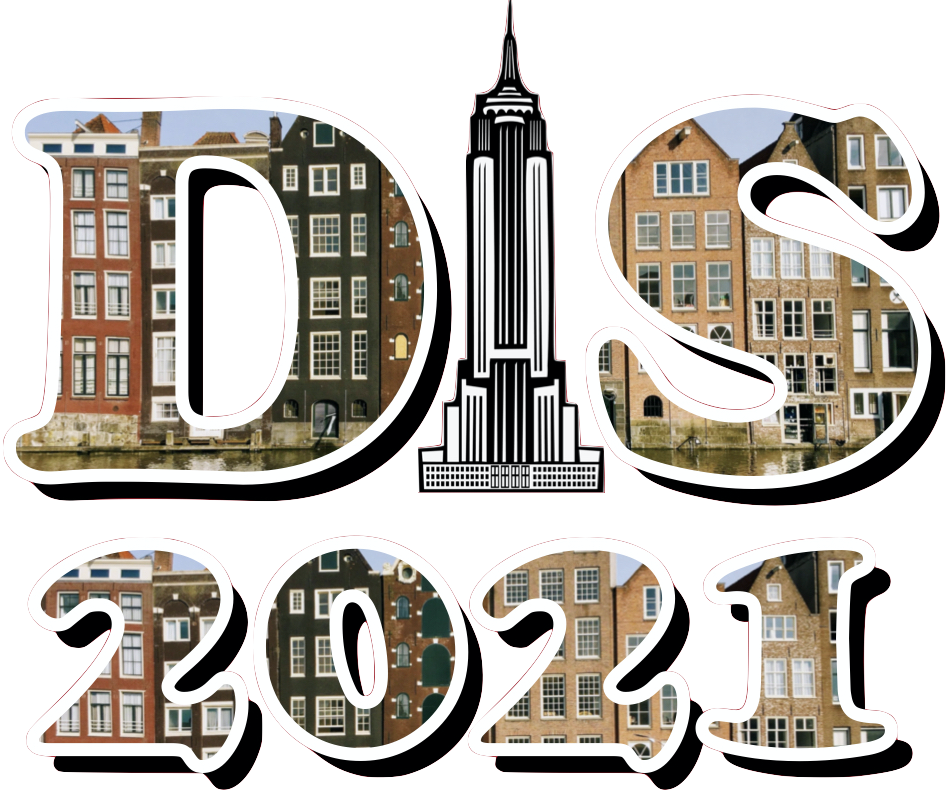}
  \end{minipage}
  &
  \begin{minipage}{0.75\textwidth}
    \begin{center}
    {\it Proceedings for the XXVIII International Workshop\\ on Deep-Inelastic Scattering and
Related Subjects,}\\
    {\it Stony Brook University, New York, USA, 12-16 April 2021} \\
    \doi{10.21468/SciPostPhysProc.?}\\
    \end{center}
  \end{minipage}
\end{tabular}
}
\end{center}

\section*{Abstract}
{\bf
% TODO: write your abstract here.
We propose the inclusive hadroproduction of a heavy-light dijet system, as a new channel for the investigation of high energy QCD. We build up a hybrid factorization that incorporates a partial next-to-leading BFKL resummation inside the standard collinear description of observables. We present a detailed analysis of different distributions: cross-section summed over azimuthal angles and differential in rapidity and azimuthal distribution. The stability that these distributions show under higher-order corrections motivates our interest in future studies, doable at new generation colliding machines. 
}

\vspace{10pt}
\noindent\rule{\textwidth}{1pt}
\tableofcontents\thispagestyle{fancy}
\noindent\rule{\textwidth}{1pt}
\vspace{10pt}

\section{Introduction}
\label{sec:intro}
% TODO: write your article here.
Semihard processes ($s \gg Q^2 \gg \Lambda_{{\rm{QCD}}}^2$, with $s$ the squared center-of-mass energy, $Q^2$ the process
hard scale and $\Lambda_{{\rm{QCD}}}^2$ the QCD mass scale) represent a challenge for high-energy QCD. Fixed-order perturbative calculations miss the effect of large energy logarithms, which must be resummed to all orders. The most general framework to perform this resummation is the BFKL approach \cite{Fadin:1975cb, Kuraev:1977fs, Balitsky:1978ic}. This procedure
allows us to resum all terms proportional to $(\alpha_s \ln(s))^n$ (the so called leading logarithmic approximation or LLA) and also of those proportional to $\alpha_s (\alpha_s \ln(s))^n$ (the so called
next-to-leading logarithmic approximation or NLA).

Within the BFKL approach, the imaginary part of an elastic amplitude (which can be related to a inelastic total cross-section by the \textit{optical theorem}), can be presented as a convolution of a process-independent object, the BFKL \textit{Green's function}, and of two process-dependent \textit{impact factors} (depending on the type of colliding particles). Being the Green's function universal (and known with NLA accuracy), to investigate new processes within the framework, all that remains is to compute impact factors. \color{black}   

In the BFKL treatment, the analytic form of the imaginary part of the process scattering amplitude is expressed as a convolution of two \emph{impact factors}, portraying the transition from each colliding particle to the respective final-state object, and of a \emph{Green's function}. 
The latter is a process-independent object and it is ruled by an integral evolution equation.
Impact factors, instead, depend on the process, thus representing the most challenging part of the calculation. A detailed list of impact factors known to the leading order (LO) and to the next-to-leading order (NLO) can be found in Ref. \cite{Celiberto:2020wpk}. In the present work the LO impact factor for the production of a forward jet (see e.g. \cite{Caporale:2014gpa} and references therein) and the one for the production of a heavy quark-antiquark pair starting from a gluon will come into play \cite{Bolognino:2019yls}.
Here, the semi-hard reaction under investigation is:
\begin{equation}
 \label{process}
 P(k_{1}) \, + \, P(k_{2}) \; \to \; Q{\text{-jet}}(k_Q, y_Q) \, + \, X \, + \, {\rm{jet}}(k_J, y_J) \; ,
\end{equation}
where $P$ labels the colliding protons with momenta $k_{i}$ ($i=1,2$), $Q$ the heavy-jet carrying (a large) transverse momenta $k_Q$ and emitted at rapidity $y_Q$ and, lastly, $J$ is the light-jet carrying (a large) transverse momenta $k_J$ and emitted at rapidity $y_J$. $X$ contains all the undetected products of the reaction.

\color{black}

\section{Theoretical set-up}
For the process under consideration we plan to construct a cross section, differential in some of the kinematic variables of the tagged jets. We will be able to build up all considered observables from the master formula for this cross-section. In the BFKL approach the partonic cross section takes a factorized form given
by the convolution of two impact factors with the BFKL Green’s function. The formula for the differential hadronic cross-section is then established by performing the convolution with collinear parton distribution functions (PDFs).  

At LO and in the $x$-range relevant for this process at LHC ($x \sim$ $10^{-2}$ to $10^{-4}$), the dominant mechanism of production of an heavy-quark pair is the splitting of a partonic gluon (with the $Q$-jet generated either from the quark or from the antiquark); the LO \textit{heavy-quark pair} impact factor (upper vertex of our hybrid-factorized cross section) and its projection onto the BFKL eigenfunctions have been evaluated in Ref. \cite{Bolognino:2019yls}. \\
Conversely, a light jet directly stems from a partonic light quark or from a gluon; the LO expression for the \textit{forward-jet} impact factor (lower-vertex of our hybrid-factorized cross section) and its projection, can be found, for instance, in Ref.\cite{Caporale:2013uva}. At this level of accuracy, the jet selection functions are obviously set to one, while impact factors are convoluted with the PDFs to obtain the hadronic cross section.
The differential proton cross section is then
\begin{equation}
\label{crosfin}
\frac{d\sigma_{pp}}{d y_Q d y_J d |\vec{k}_Q| d |\vec{k}_J| d \phi_Q d \phi_J}=\frac{1}{(2\pi)^2} \left[\mathcal{C}_0+2 \sum_{n=1}^{\infty} \cos(n\phi) \mathcal{C}_n \right], 
\end{equation}
where $\varphi = \phi_J - \phi_Q - \pi$ and $\mathcal{C}_n$'s are the \textit{azimuthal coefficients}.  $\mathcal{C}_n$'s embody in their definition the parton distribution function, impact factors and the BFKL Green function (projected on the $(n, \nu)$-space). Their analytic expression can be found in Ref. \cite{Bolognino:2021mrc}. \color{black}

\section{Phenomenology}
\label{sec:Pheno}
Observables under investigation are:
\begin{itemize}

    \item \emph{Azimuthal-angle coefficients}
    \begin{equation}
     \label{Cn_int}
     C_n(\Delta Y, s) =
     \int_{k^{\rm min}_Q}^{k^{\rm max}_Q} d |\vec k_Q|
     \int_{k^{\rm min}_J}^{{k^{\rm max}_J}} d |\vec k_J|
     \int_{y^{\rm min}_Q}^{y^{\rm max}_Q} d y_Q
     \int_{y^{\rm min}_J}^{y^{\rm max}_J} d y_J
     \, \delta \left( y_Q - y_J - \Delta Y \right)
     \, {\cal C}_n %\left(|\vec p_H|, |\vec p_J|, y_H, y_J \right)
     \, ,
    \end{equation}
    We investigate the $\Delta Y$-behavior of the $\varphi$-summed cross section (or $\Delta Y$-distribution), $C_0(\Delta Y, s)$.
    \item \emph{Azimuthal distribution} 
    \begin{equation}
     \label{azimuthal_distribution}
%     \frac{1}{\sigma_{pp}} 
     \frac{d \sigma_{pp} (\varphi, \Delta Y, s)}{\sigma_{pp} d \varphi} \equiv \frac{1}{\pi} \left\{ \frac{1}{2} + \sum_{n=1}^{\infty} \cos(n \varphi) \frac{C_{n}}{C_0} \right\} = \frac{1}{\pi} \left\{ \frac{1}{2} + \sum_{n=1}^{\infty} \cos(n \varphi) \langle \cos(n \varphi) \rangle \right\}
      \; .
    \end{equation}
    \end{itemize}
In the following, we focus our attention on bottom-jet ($b\text{-jets}$) emissions fixing the center-of-mass energy at $\sqrt{s} = 14$ TeV. Rapidity intervals, $|y_Q| < 2.5$ and $|y_J| < 4.7$, are choosen in order to fit a realistic CMS analysis. We work in the ${\overline {\rm MS}}$ scheme. \\ \color{black}
An extensive analysis, which includes other distributions (\textit{e.g.} the heavy-jet transverse momentum distribution) as well as the charm-jet production case, can be found in \cite{Bolognino:2021mrc}.

\begin{figure}
\centering
\includegraphics[scale=0.48,clip]{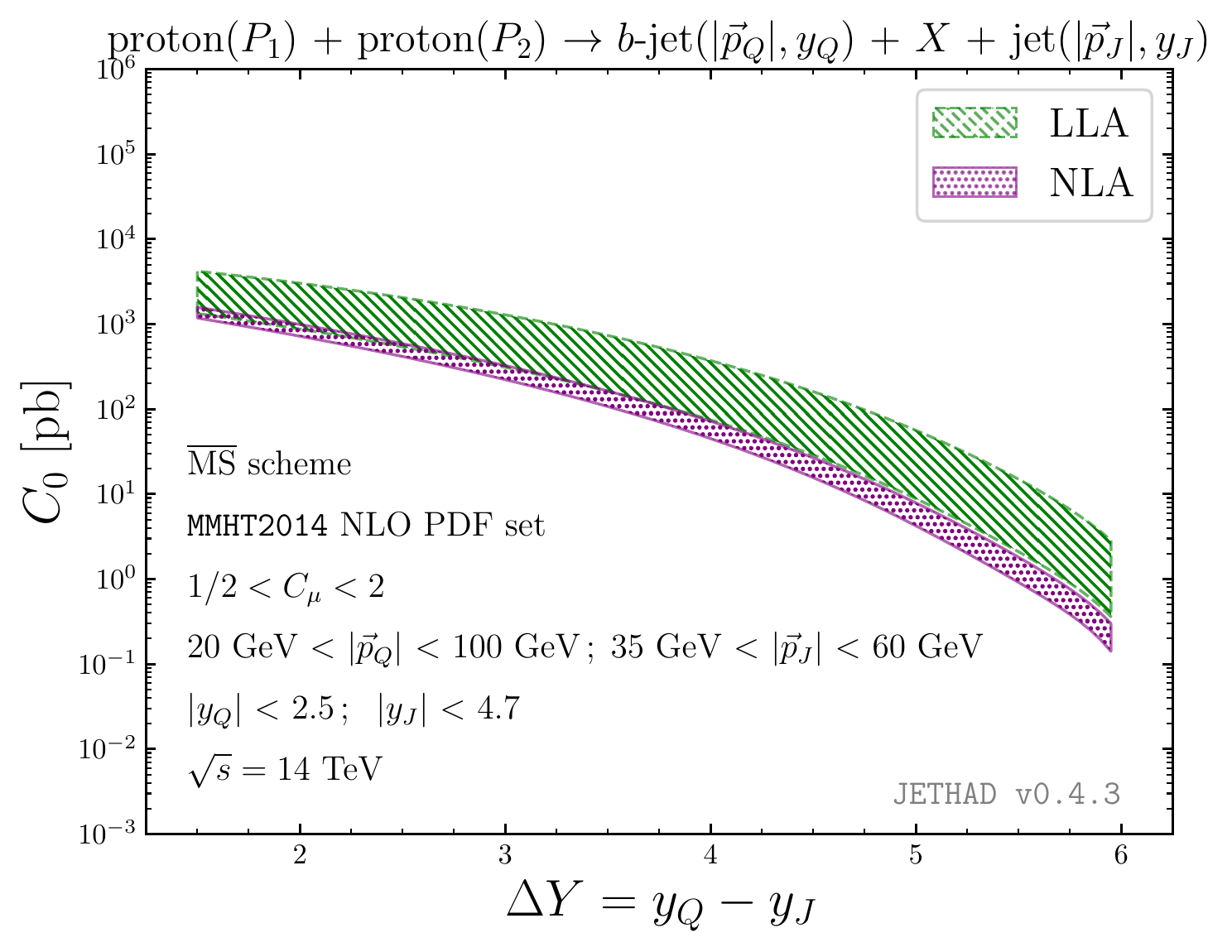}
\includegraphics[scale=0.48,clip]{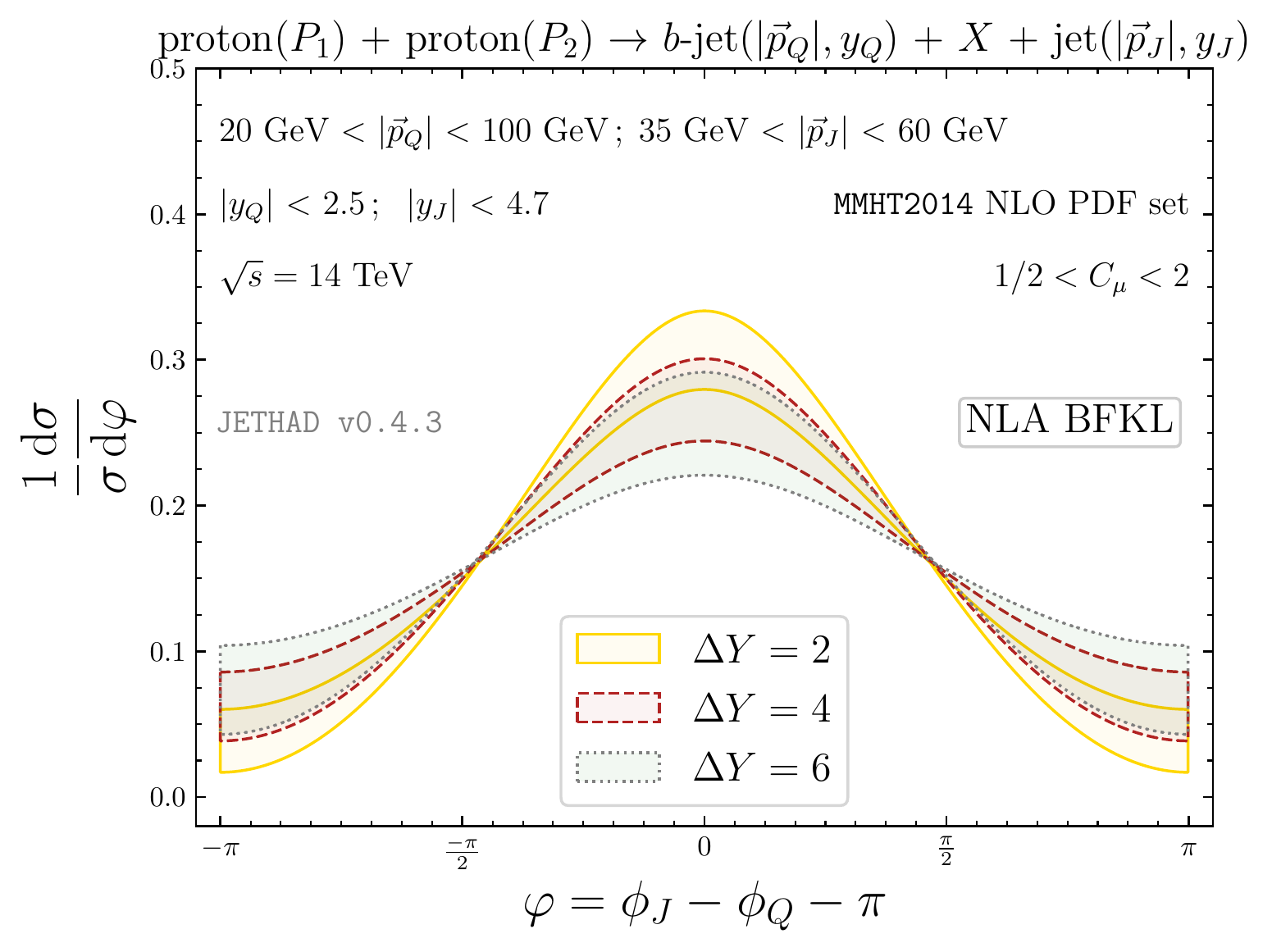}
\caption{$\varphi$-summed cross section, $C_0$ as a function of $\Delta Y$ (left). NLA azimuthal distribution for different $\Delta Y$ (right).}
\label{fig:C0}
\end{figure}
In the left panel of Fig. \ref{fig:C0} we present predictions for the $\varphi$-summed cross section, $C_0$, as a function of the rapidity interval, $\Delta Y$. The downtrend of the cross section comes out as the net result of two competing effects. On one side BFKL predicts an increase in the partonic cross section as the rapidity grows. On the other side, in our hybrid approach the BFKL-resummed partonic cross section is convoluted with PDFs. In particular, the gluon one plays a dominant role and it decreases as the rapidity difference grows. We note that the inclusion of NLA corrections to the BFKL Green's function reduces the value of the cross section, due to the fact that these terms have opposite sign with respect to pure LLA. Nevertheless, the emission of a heavy-flavored jet provides with a partial stabilization, that allows us to perform studies around small multiples of natural scales.  

In the right panel of Fig. \ref{fig:C0} we present predictions for the azimuthal distribution, $d \sigma_{pp}/d\varphi$ at different values of $\Delta Y$. High-energy behavior clearly emerges from these distributions. In fact, all distributions have a peak when the two tagged objects are back-to-back ($\varphi=0$), but, distributions associated to higher rapidity difference are wider and with a smaller peak height, indicating a clear loss of correlation when $\Delta Y$ increases. \color{black}

\section{Conclusions and outlook}
We have proposed a new type of inclusive semi-hard reaction in which a heavy-flavored jet and a light-flavored jet are emitted with a large separation in rapidity. 
The following phenomenological analysis, not only enriches the class of useful processes for testing the high-energy dynamics of QCD (see \textit{e.g.} \cite{Bolognino:2018oth,Celiberto:2020tmb,Celiberto:2021dzy,Celiberto:2021fdp,Caporale:2015vya,Caporale:2015int}), but also allows for a stabilization of the BFKL series. Thanks to these stabilizing effects, it has been possible, in contrast with the case of inclusive light-dijet production ~\cite{Colferai:2010wu,Caporale:2012ih,Ducloue:2013hia,Ducloue:2013bva,Caporale:2013uva,Caporale:2014gpa,Colferai:2015zfa,Caporale:2015uva,Ducloue:2015jba,Celiberto:2015yba,Celiberto:2015mpa,Celiberto:2016ygs,Celiberto:2016vva,Caporale:2018qnm} (Mueller-Navelet channel~\cite{Mueller:1986ey}), to perform studies around natural values of energy scales. This result is very promising when considering the perspective of performing high-precision calculations in our hybrid collinear/high-energy framework, as well as gauging the weight of high-energy effects in wider kinematical ranges, like the ones typical of new generation colliding machines \cite{AbdulKhalek:2021gbh,Chapon:2020heu,Arbuzov:2020cqg,Anchordoqui:2021ghd}. 

This work is part of our current program on heavy-flavored emissions, which has started from the analytic calculation of heavy-quark pair impact factors \cite{Celiberto:2017nyx,Bolognino:2019ouc,Bolognino:2019yls}. From the phenomenological point of view, our next step is the extension of these analyses to bound states of quarks, such as heavy-light mesons and quarkonia. From the theoretical side a very interesting and challenging development is the calculation of the NLO heavy-quark pair impact factor. 

% TODO: include author contributions

% TODO: include funding information

% TODO:
% Provide your bibliography here. You have two options:

% FIRST OPTION - write your entries here directly, following the example below, including Author(s), Title, Journal Ref. with year in parentheses at the end, followed by the DOI number.
%\begin{thebibliography}{99}
%\bibitem{1931_Bethe_ZP_71} H. A. Bethe, {\it Zur Theorie der Metalle. i. Eigenwerte und Eigenfunktionen der linearen Atomkette}, Zeit. f{\"u}r Phys. {\bf 71}, 205 (1931), \doi{10.1007\%2FBF01341708}.
%\bibitem{arXiv:1108.2700} P. Ginsparg, {\it It was twenty years ago today... }, \url{http://arxiv.org/abs/1108.2700}.
%\end{thebibliography}

% \bibliographystyle{SciPost_bibstyle} % Include this style file here only if you are not using our template
\bibliography{SciPost_LaTeX_Template.bib}

\nolinenumbers

\end{document}